\documentclass[aps, prb, reprint, lengthcheck, footinbib, showpacs, superscriptaddress, citeautoscript]{revtex4-1}

\usepackage[T1]{fontenc}
\usepackage[english]{babel}

\usepackage{amsmath}

\usepackage{braket}
\usepackage{graphicx}

\usepackage[colorlinks, urlcolor=blue, citecolor=blue, linkcolor=blue, pdfstartview=FitH]{hyperref}
\usepackage[all]{hypcap}

\begin{document}

\def\affiSPBU{Department of Solid State Physics, Saint~Petersburg State University, 198504 St.~Petersburg, Russia}
\def\affiSOLAB{Spin Optics Laboratory, Saint~Petersburg State University, 198504 St.~Petersburg, Russia}
\def\affiE2a{Experimentelle Physik 2, Technische Universit\"at Dortmund, D-44221 Dortmund, Germany}
\def\affiIOFFE{Ioffe Physical-Technical Institute, Russian Academy of Sciences, 194021 St.~Petersburg, Russia}
\def\affiPad{Department Physik, Universit\"at Paderborn, 33098 Paderborn, Germany}
\def\affiBochum{Angewandte Festk\"orperphysik, Ruhr-Universit\"at Bochum, D-44780 Bochum, Germany}

\title{Spin dynamics of quadrupole nuclei in InGaAs quantum dots}

\author{M.~S.~Kuznetsova}
\affiliation{\affiSPBU}
\author{R.~V.~Cherbunin}
\affiliation{\affiSPBU}
\author{I.~Ya.~Gerlovin}
\affiliation{\affiSOLAB}
\author{I.~V.~Ignatiev}
\affiliation{\affiSPBU}
\affiliation{\affiSOLAB}
\author{S.~Yu.~Verbin}
\affiliation{\affiSPBU}
\author{D.~R.~Yakovlev}
\affiliation{\affiE2a}
\affiliation{\affiIOFFE}
\author{D.~Reuter}
\affiliation{\affiPad}
\author{A.~D.~Wieck}
\affiliation{\affiBochum}
\author{M.~Bayer}
\affiliation{\affiE2a}
\affiliation{\affiIOFFE}

\date{\today}

\begin{abstract}
Photoluminescence polarization is experimentally studied for samples with  (In,Ga)As/GaAs self-assembled quantum dots in transverse magnetic field (Hanle effect) under slow modulation of the excitation light polarization from fractions of Hz to tens of kHz. The polarization reflects the evolution of strongly coupled electron-nuclear spin system in the quantum dots. Strong modification of the Hanle curves under variation of the modulation period is attributed to the peculiarities of the spin dynamics of quadrupole nuclei, which states are split due to deformation of the crystal lattice in the quantum dots. Analysis of the Hanle curves is fulfilled in the framework of a phenomenological model considering a separate dynamics of a nuclear field $B_{Nd}$ determined by the $\pm 1/2$ nuclear spin states and of a nuclear field  $B_{Nq}$ determined by the split-off states $\pm 3/2$, $\pm 5/2$, etc. It is found that the characteristic relaxation time for the nuclear field $B_{Nd}$ is of order of 0.5~s, while the relaxation of the field $B_{Nq}$ is faster by three orders of magnitude. 
\end{abstract}


\maketitle

\section*{Introduction}
Hyperfine interaction of an electron localized in a quantum dot (QD) with nuclear spins forms a strongly coupled electron-nuclear spin system~\cite{OO, Review-Kalevich}. This system is considered to be promising for realization of quantum information processing devices~\cite{KaneNature98, TaylorPRL03, BoehmeSc12}. The realization of spin qubits assumes some stability of the spin system required for the storage and processing of quantum information. In QDs, the optically polarized electron transfers its spin moment into the nuclear subsystem where the spin orientation may be conserved for a long time controlled by the nuclear spin relaxation processes. 

The main process destroying the nuclear spin polarization is believed to be the transverse relaxation in local fields caused by the dipole-dipole interaction of neighboring nuclear spins. Characteristic time of the relaxation, $T_2$, for nuclei with spins $I = 1/2$ is of order of 10$^{-4}$~s~\cite{OO}. The effective local fields, $B_{dd}$, are of fraction of milliTesla and can be easily suppressed by external magnetic fields exceeding these local fields. 

In the case of self-assembled QDs, the stabilization of nuclear spin orientation is possible, in principle, in the absence of external magnetic field~\cite{OultonPRL07}. Due to noticeable difference in the lattice constants of QDs and barrier layers, some elastic stress appears in the QDs causing mechanical deformation of crystal lattice. The deformation results in a gradient of crystal fields acting on nuclei from neighboring atoms and splitting the nuclear spin states for quadrupole nuclei with $I > 1/2$~\cite{OO}. Because the strain-induced quadrupole splitting in self-assembled QDs typically greatly exceeds Zeeman splitting in the local fields, the spin orientation of quadrupole nuclei is pinned to the principal deformation axis and is not destroyed by the dipole-dipole interaction~\cite{DzhioevPRL07}. In this case, the stability of nuclear spin system should be determined by processes of longitudinal spin relaxation of quadrupole nuclei with characteristic time $T_1 >> T_2$. Although many publications are devoted to the nuclear spin polarization~\cite{GammonPRL01, Kroutvar-Nature2004, IkezawaPRB05, MaletinskyPRL07, OultonPRL07, AkimovPRL06, BraunPRB06, EblePRB06, LaiPRL06, TartakovskiiPRL07, BelhadjPRB08, KajiPRB08, Skiba-SzymanskaPRB08, KrebsPRL10, ChekhovichPRL10, ChekhovichNature12, HoglePRL12, VerbinJETF12, KuznetsovaPRB13, CherbuninPRB09} (see also review articles~\cite{ Review-Kalevich, CoishPSSB09, ChekhovichNatureMat12, UrbaszekRMP13}), there are very few works where relaxation dynamics is studied for quadrupole nuclei~\cite{DengPRB05, PagetPRB08, Kotur16}.

In this paper we report on experimental study of spin dynamics of quadrupole nuclei in the singly-charged (In,Ga)As/GaAs QDs. The nuclear spin polarization was studied in optical experiments by detection of the electron spin orientation via polarized secondary emission of the QDs in a transverse magnetic field (the Hanle effect).
We have found that, when the photoluminescence (PL) of the samples under study is excited by light with the modulated helicity of polarization, the Hanle curves strongly depend on the modulation frequency. 
We have developed a phenomenological model based on the consideration of separate polarization dynamics of the $|\pm 1/2>$ nuclear spin doublets and of the split-off doublets, $|\pm 3/2>$, $|\pm 3/2>$, etc. The analysis performed using a pseudo-spin
approach proposed in  Ref.~\onlinecite{ArtemovaFTT85} has allowed us to extract contributions from polarization of these different groups of spin doublets into the effective nuclear field acting on the electron spin.

\section{Experimental details\label{sec:Experimental details}}

We studied two samples prepared from one heterostructure with InAs/GaAs QDs grown by Stranski-Krastanov method. Sample A was then annealed at temperature $900$~$^{\circ}$C and sample B at temperature $980$~$^{\circ}$C. The annealing gives rise to the diffusion of indium atoms into the barriers so that the indium concentration and, correspondingly, the crystal lattice deformations decrease for higher annealing temperature. Theoretical modeling shows~\cite{PetrovPRB08} that the deformation is of about 3 $\%$ for sample A and 1 $\%$ for sample B. The quadrupole splitting of the nuclear spin states strongly depends on the deformation~\cite{SokolovPRB16}. It is considerably smaller for sample B comparing to sample A. Therefore, the experimental study and analysis of two samples allows one to highlight the role of quadrupole splitting of nuclear states in the observed effects.

The QDs under study contain one resident electron per dot on average due to $\delta$-doping of barriers by donors during the epilaxial growth. There are 20 layers of the QDs with areal density of about $10^{10}$ cm$^{-2}$ separated by 60-nm thick GaAs barriers~\cite{GreilichPRL06}. Optical characterization of the samples is given in Ref.~\onlinecite{KuznetsovaPRB14}. The photoluminescence (PL) band in sample A corresponding to the lowest optical transitions in the QDs is centered at photon energy $E_A = 1.34$~eV with the half width at half maximum (HWHM), $\delta E_A = 9$~meV. Similar PL band in sample B is shifted to the higher photon energy due to smaller indium content, $E_B = 1.42$~eV with $\delta E_B = 7$~meV. 

In our present experiments, dependence of circular polarization of PL is measured as a function of the magnetic field applied perpendicular to the optical axis. The depolarization curves (Hanle curves) are measured under optical excitation by a continuous wave Ti:sapphire laser into the wetting layer of each sample ($E_{WL} = 1.459$~eV for sample A and $E_{WL} = 1.481$~eV for sample B). Polarization of the laser radiation is slowly modulated between $\sigma^{+}$ and $\sigma^{-}$ by an electro-optical modulator followed by a quarter-wave plate with a frequency varied from fractions of Hz to several kHz. No resonant effects studied in Refs.~\onlinecite{FlisinskiPRB10, KuznetsovaPRB14} are observed at such slow modulation of the polarization. 

The PL is dispersed by a 0.5-m spectrometer and detected with a silicon avalanche photodiode. The circular polarization degree, $\rho = (I^{++} - I^{+-}) / (I^{++} + I^{+-})$, is measured using a photo-elastic modulator operating at a frequency of 50~kHz and a two-channel photon counting system. Here $I^{++}$($I^{+-}$) is the PL intensity for co- (cross-) circular polarization relative to that of excitation. In the maximum of PL band of the QDs, the polarization is negative and reflects the mean spin polarization of resident electrons as it was extensively discussed earlier~\cite{IgnatievOS09, CherbuninPRB11}. Hereafter we use the maximal absolute value of $\rho$ obtained at the center of PL band for each sample, $A_{NCP} = \max |\rho(\omega)|$, for quantitative characteristic of the electron spin polarization~\cite{IgnatievOS09}: $S_z = A_{NCP}/2$, along the optical axis. Because the resident electrons are interacting with the QD nuclei, the negative circular polarization (NCP) can be used as a sensitive tool to monitor the nuclear spin state~\cite{IgnatievOS09, CortezPRL02, ShabaevPRB09}.

\section{Experimental results and analysis\label{sec:II}}

\subsection{Hanle curves at optical excitation with modulated polarization.}

\begin{figure*}[t]
\includegraphics[width=2\columnwidth,clip]{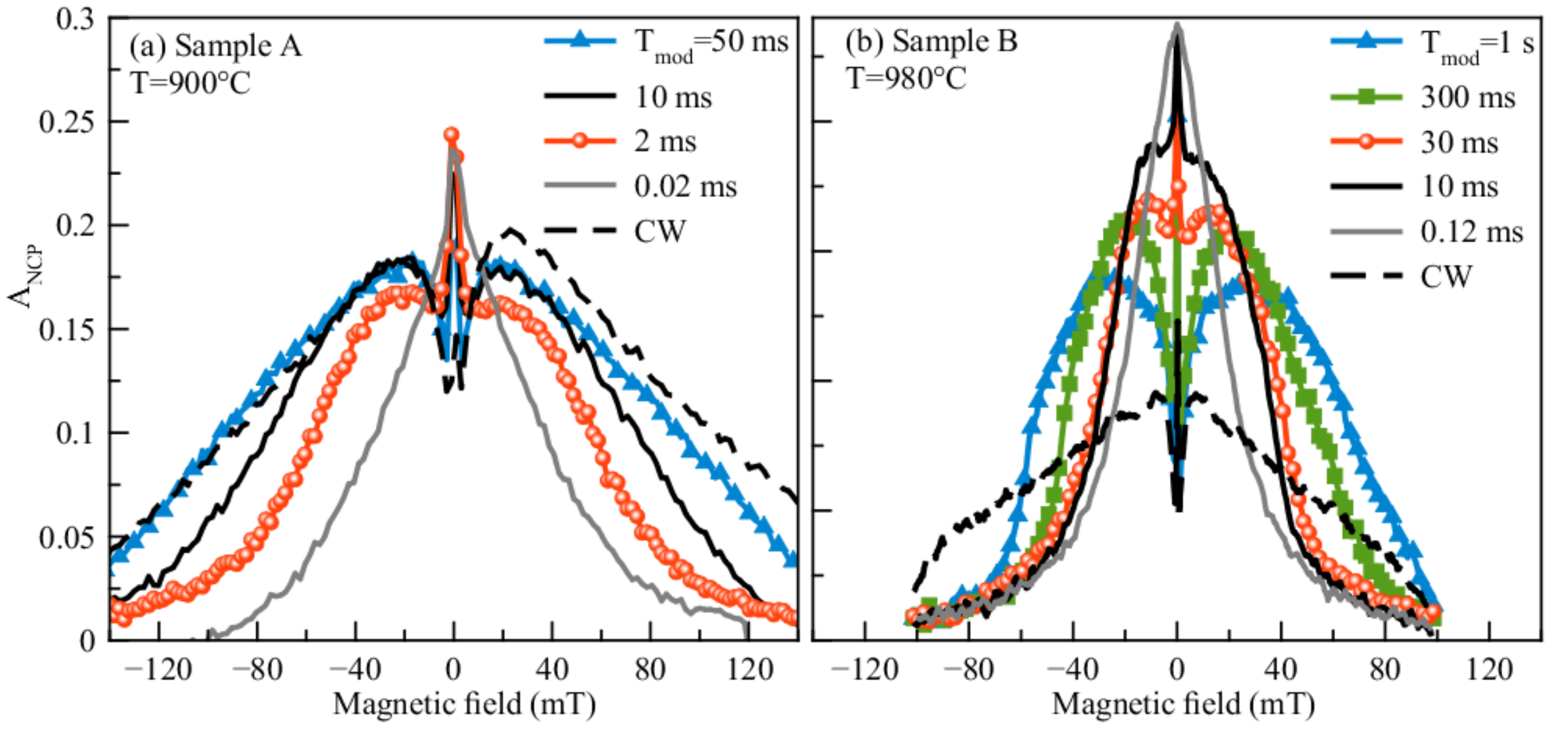}
\caption{(Color online) Hanle curves for sample A (left panel) and sample B (right panel) measured at the optical excitation of one circular polarization (CW) as well as with modulated polarization with periods given in the legends. Excitation power $P = 14$~\textrm{mW} for sample A and 10~mW for sample B. Diameter of laser spots on the samples, $d \approx 60$~$\mu$m. $T = 1.8$~K.%
}\label{fig:Fig1}
\end{figure*}

Typical Hanle curves for different modulation periods of excitation polarization are shown in Fig.~\ref{fig:Fig1}. As one can see, the Hanle curves of sample A [panel (a)] annealed at lower temperature is considerably broader than the curves of sample B [panel (b)]. For both samples, the shape of Hanle curves strongly depends on the modulation period. At large periods, as well as at the excitation with a fixed polarization, a well resolved W-structure in the small magnetic fields is observed indicating a dynamic nuclear polarization (DNP) acting on electron spin as an effective nuclear field ~\cite{PagetPRB77}. The W-structure becomes smoothed and then almost disappears when the modulation period decreases. Besides, the Hanle curves noticeably shrink with the period shortening. At the smallest periods used in the experiments, the Hanle curves acquire almost Lorentzian shape. Experiments also show that the Hanle curve width becomes independent on the modulation frequency at its further increase but monotonically increases with the excitation power (not shown here).
Such regularity is typical for depolarization of electron spins with no nuclear spin effects~\cite{OO}.

It is important that, in spite of the large overall modification of Hanle curves, the polarization degree measured at zero magnetic fields is almost independent of modulation frequency. It approaches some value with the rise of excitation power and becomes also almost independent of the power at strong enough excitation. 
We assume that this stability of polarization indicates the total polarization of electron spins in the QDs at zero magnetic field. The deviation of the experimentally obtained value of the polarization from unity is most probably caused by contribution of non-polarized PL from the neutral or doubly-charged QDs~\cite{KuznetsovaPRB14}.

Remarkable difference in behavior of Hanle curves is observed for two samples studied, compare Fig~\ref{fig:Fig1}(a) and Fig~\ref{fig:Fig1}(a). Namely, for sample B with higher annealing temperature, strong modification of Hanle curve is observed even at large modulation period, $T_{mod} = 1$~s, while for sample A the modification is hardly seen at the ten times shorter period. Besides, the Hanle curve narrowing for sample B is followed by strong increase of polarization far beyond the W-structure. No such increase is observed for sample A. This difference in the Hanle curve behavior indicates large difference in the dynamics of nuclear spin system in these two samples.

The analysis of complex shape of the Hanle curves is the main topic of the rest part of the paper. As it is shown in Ref.~\onlinecite{KuznetsovaPRB13}, the W-structure and the shape of central part of the Hanle curves for the QDs under study can be well described in the framework of a phenomenological model. The model considers the electron spin precession about an effective magnetic field, which is the sum of the external magnetic field, $\mathbf{B}$, an effective field of the DNP (Overhauser field)~\cite{OverhauserPR53}, $\mathbf{B}_{N}$, and an effective field of the nuclear spin fluctuations, $\mathbf{B}_{f}$~\cite{MerkulovPRB02}.

In the GaAs-based structures with no quadrupole effects, the regular nuclear field is developed, in Hanle experiments, parallel, rather than anti-parallel, to the external magnetic field because of the negative sign of electron $g$ factor~\cite{OO}. The W-structure, in particular, the dips in the structure, are formed due to the large nuclear field, which magnifies the effect of external magnetic field on the electron spin~\cite{PagetPRB77, KuznetsovaPRB13}. At larger magnetic fields, i.e., at the wings of Hanle curves, the electron polarization is additionally suppressed by the nuclear field added to the external magnetic field. Therefore, it would be expected that the modulation of excitation modulation suppressing the nuclear polarization should partially restore the electron polarization at the wings of Hanle curve. 

Experimentally observed evolution of the Hanle curves strongly differs from this prediction. As it is seen in~Fig.~\ref{fig:Fig1}, the increase of modulation frequency is followed by a smoothing of the W-structure that indicates the decrease of nuclear polarization. At the same time, the width of Hanle curves decreases, rather than increases, at it is predicted by the standard model~\cite{OO}. 

We assume that the main reason for such behavior is the quadrupole effects in the nuclear spin system~\cite{DzhioevPRL07}. A gradient of crystal field splits off the spin doublets $|\pm 3/2>$, $|\pm 5/2>$, etc, from doublet $|\pm 1/2>$. In the structures under study, the gradient is mainly induced by the crystal lattice deformation. The principal axis of this deformation is directed along the growth axis of the structures that is along the optical axis in our experiments~\cite{SokolovPRB16}. The quadrupole splitting caused by this deformation is studied in detail in Ref.~\onlinecite{KuznetsovaPRB14}. 

At the presence of quadrupole splitting, behavior of the $|\pm 1/2>$ doublet and the split-off doublets in the magnetic field orthogonal to deformation axis (transverse field) is very different. Hereafter we call these components as the dipole and quadrupole components of the nuclear field, respectively. 

 According to  Ref.~\onlinecite{DzhioevPRL07} the split-off nuclear states are not splitted in the transverse magnetic field at the first order. This means that the quadrupole components of nuclear field conserve their orientation. Correspondingly, the electron spin polarization is also conserved due to hyperfine interaction with the stabilized nuclear spins. Only when the Zeeman splitting becomes comparable with the quadrupole splitting, the nuclear spins are no longer pinned to the major axis of the electric field gradient. Correspondingly, the nuclear spin orientation is destroyed and electron polarization decreases. That forms the wings of the Hanle curve. The fast modulation of excitation suppressing the quadrupole component of nuclear field should result in destroying the nuclear polarization at smaller magnetic fields that is in narrowing the Hanle curve.

This simplified discussion of dynamic processes in the electron-nuclear spin system allows one to qualitatively explain the experimentally observed behavior of Hanle curves at different modulation frequencies.
An accurate analysis of the Hanle curves allowed us to obtain valuable information about both the dipole and quadrupole components of nuclear field.

\subsection{Phenomenological model}

To extract information about the dynamics of nuclear polarization from the Hanle curves, we generalize the phenomenological model proposed in Ref.~\onlinecite{KuznetsovaPRB13}. In particular, we consider two effective nuclear fields acting on the electron spin. The first one, the dipole field $\mathbf{B}_{Nd}$, is determined by polarization of the $\pm 1/2$ nuclear spin states. The second one, the quadrupole field $\mathbf{B}_{Nq}$, is due to the polarization of the split-off states $\pm 3/2$, etc. We should note that, in (In,Ga)As-based structures, the nuclei of all chemical elements, including isotopes constituting the structure, possess quadrupole moments. 

The electron spin precesses in the total field, $\mathbf{B}_\mathrm{tot}$, consisting of several contributions:
\begin{equation}
   \mathbf{B}_\mathrm{tot}=\mathbf{B}+\mathbf{B}_{Nd}+\mathbf{B}_{Nq}+\mathbf{B}_{f},
\label{eq:Eq3}
\end{equation}
where $\mathbf{B}_f$ is an effective field of the nuclear spin fluctuations.
 Due to the fast precession of electron spin about $\mathbf{B}_\mathrm{tot}$, only the projection, $S_{B_\mathrm{tot}}$, is conserved:
\begin{equation}
S_{{B}_{\mathrm{tot}}}=\frac{{\mathbf{(S_{0}\cdot B}_\mathrm{tot})}}{\lvert\mathbf{B}_\mathrm{tot}\rvert}=S_{0}\frac{B_{\mathrm{tot}\emph{z}}}{\sqrt{B^2_\mathrm{tot}}}.
\label{eq:Eq4}
\end{equation}
Here ${S_{0}}$ is the electron spin polarization created along the optical axis ($z$-axis). The electron spin polarization measured in the experiments $S_z$ is the projection of ${S_{B_{tot}}}$ on the direction of observation (the optical axis). Correspondingly, the measured degree of PL polarization is:
\begin{equation}
\rho=\frac{S_z}{S_{0}}=\frac{B^2_{\mathrm{tot}\emph{z}}}{B^2_\mathrm{tot}} .
\label{eq:Eq5}
\end{equation}
The electron spin precession competes with the spin relaxation, which can be described by an effective field $B_{\tau} = \hbar/(g_e\mu_B\tau_s)$ where $g_e$ is the electron $g$ factor, $\mu_B$ is the Bohr magneton, and $\tau_e$ is the electron spin relaxation time.
To include the relaxation, we should generalize Eq.~\eqref{eq:Eq5}:
\begin{equation}
\rho=\frac{B^2_{\mathrm{tot}\emph{z}}+B^2_{\tau}}{B^2_\mathrm{tot}+B^2_{\tau}} .
\label{eq:Eq6}
\end{equation}
For simplicity, we assume here that the relaxation time $\tau_{e}$ does not depend on the external magnetic field. This assumption will be verified by the simulations of Hanle curves described below. Similar to Ref.~\onlinecite{KuznetsovaPRB13}, we assume that the total field squared can be expressed as:
\begin{equation}
B^2_\mathrm{tot}= (B+B_{Ndx}+B_{Nqx})^2+ (B_{Ndz}+B_{Nqz})^2+\braket {B^2_{f}}.
\label{eq:Eq7}
\end{equation}
Here we use the fact that the external magnetic field is directed along the $x$-axis. We also assume that no valuable nuclear polarization appears along the $y$-axis. The nuclear spin fluctuations are assumed to be isotropically distributed:
\begin{equation}
\braket{B^2_{f}}=\braket{B^2_{fx}}+\braket{B^2_{fy}}+\braket{B^2_{fz}}=3\braket{B^2_{fz}}.
\label{eq:Eq8}
\end{equation}
The $z$-projection of the total field squared, $B^2_{\mathrm{tot}\emph{z}}$, is determined by similar way with taking into account only $z$-components of the regular and fluctuating fields. 
Finally we obtain:
\begin{widetext}
\begin{equation}
\rho(B)=\frac{B_e}{B_e^0}=\frac{(B_{Ndz}+B_{Nqz})^2+\braket{B^2_{fz}}+B^2_{\tau}}{(B+B_{Ndx}+B_{Nqx})^2+(B_{Ndz}+B_{Nqz})^2+3\braket{B^2_{fz}}+B^2_{\tau}}.
\label{eq:Eq9}
\end{equation}
\end{widetext}
Here $B_e = b_e S_z$ is the $z$-component of Knight field acting on the nuclei and  $B_e^0 = b_e S_0$  is the Knight field at zero external magnetic field. Constant $b_e$ is proportional to the hyperfine interaction constant~\cite{OO}. It is considered as a fitting parameter. 

We suppose that components of the nuclear field, $B_{Ndx}$ and $B_{Ndz}$, $B_{Nqx}$ and $B_{Nqz}$ are determined by the nuclear spin precession about the total field acting on the nuclei. The field consists of the external magnetic field, $B$, and of the Knight field, $B_e$. For simplicity, we neglect the $x$- and $y$-components of the Knight field because they are much smaller than the external magnetic field. 

Evolution of the nuclear field created by nuclei with quadrupole splitting of spin states can be analyzed in the framework of a pseudo-spin model proposed in Ref.~\onlinecite{ArtemovaFTT85}. According to the model, each spin doublet with the spin projection, $m$ = $\pm 1/2$, $\pm 3/2$, \ldots, onto the principal quadrupole axis may be considered independently, while the Zeeman splitting of the doublet in an external magnetic field is considerably smaller than the energy separation between the doublets determined by the quadrupole splitting. The Zeeman splitting, $\delta E_m = g_m \beta B$, can be described by an anisotropic nuclear $g$ factor, $g_m$. Here $\beta$ is the nuclear magneton. The nuclear spin polarization and, correspondingly, the nuclear field are created along an effective magnetic field, $\mathbf{B}_m^{eff}=g_{mx}\mathbf{B}+g_{mz}\mathbf{B}_e$~\cite{Note-field}.  We should stress that the direction of $\mathbf{B}_m^{eff}$  deviates, in general case, from the direction of vector sum of fields $\mathbf{B}$ and $\mathbf{B}_e$ because of the anisotropy of the nuclear $g$ factor. 

Using a simple vector model~\cite{KuznetsovaPRB13} one can obtain general expressions for components of nuclear field:
\begin{equation}
\begin{split}
    B_{N_{mz}}=B_{Nm}\frac{B^2_e}{(g_{mx}^*B)^2+B^2_e}, \\
    B_{N_{mx}}=B_{Nm}\frac{(g_{mx}^*B)B_e}{(g_{mx}^*B)^2+B^2_e},
\end{split}
\label{eq:Eq10}
\end{equation}
Here $g_{mx}^*={g_{mx}}/{g_{mz}}$ are the normalized $g$ factors determined as the ratio of $g$ factors characterizing interactions with the magnetic fields applied across and along the principal quadrupole axis, respectively. In small transverse magnetic fields, the splitting of nuclear states with $m$= $\pm 1/2$ (the dipole states) linearly depends on the magnetic field and $g_{dx}^*\approx 2$, while the Zeeman splitting of the doublet is considerably smaller than the quadrupole splitting. We will use this approximate equality because, as it will be seen in the next section, the dipole nuclear field significantly differs from zero only in small magnetic fields. 

The splitting of the $\pm 3/2$, $\pm 5/2$, \ldots, doublets is strongly anisotropic one in the transverse magnetic field and nonlinearly depends on the magnetic field magnitude. For nuclei with $I = 3/2$, splitting of the $\pm 3/2$ spin states is described by expression:~\cite{Abragam}
\begin{equation}
\delta E_{\pm3/2}=\frac{E_Q}{2}[a+(\sqrt{1-a+a^2}-\sqrt{1+a+a^2})],
\label{eq:Eq11}
\end{equation}
where $a = 4\gamma \hbar B/E_Q$. Here $E_Q$ is the quadrupole splitting of the $\pm 1/2$ and $\pm 3/2$ doublets at zero magnetic field and $\gamma$ is the gyromagnetic ratio for the nuclei. Eq.~\eqref{eq:Eq11} allows one to obtain an exact expression for the nuclear $g$ factor. We found, however, that this complex expression can be well fitted for all the nuclei and magnetic fields considered here by a phenomenological formula: 
\begin{equation}
 g_{qx}^*=k\frac{B^2}{B^2+B_\Delta^2},
\label{eq:Eq11a}
\end{equation}
where $k$ and $B_{\Delta}$ are the fitting parameters. According to this expression, the $g$ factor quadratically rises with magnetic field at small $B$ and then reaches a constant value at $B \gg B_{\Delta}$. An analysis shows that both the parameters are strongly different for Ga and As nuclei due to different quadrupole splittings. Therefore, to accurately model the nuclear field, a sum of contributions of different nuclei, like those given by Eqs.~\eqref{eq:Eq11}, would be considered. The experimental results, however, do not contain sufficient information required for separation of different contributions. We, therefore, simplify our analysis and suggest the simplest, linear, dependence for the $g$ factor,
\begin{equation}
g_{qx}^*=kB ,
\label{eq:Eq11b}
\end{equation}
to model the effective nuclear field averaged over all the nuclei. Results, described in the next subsection, show that this dependence allows us to explain main peculiarities of the Hanle curves. 

Substitution of the expressions~\eqref{eq:Eq10} into Eq.~\eqref{eq:Eq9} gives rise to an equation of the 9-th degree relative to Knight field $B_e$. Solution of this equation for different magnetic fields gives the field dependence of electron spin polarization that is the Hanle curve. Comparison of the modeled Hanle curve with that obtained experimentally allows us to determine fitting parameters $B_{\tau}$, $b_e$, $B_{Nd}$, $B_{Nq}$, $B_{fz}$, and $k$ for each modulation period.
\begin{figure}[t]
\includegraphics[width=1\columnwidth,clip]{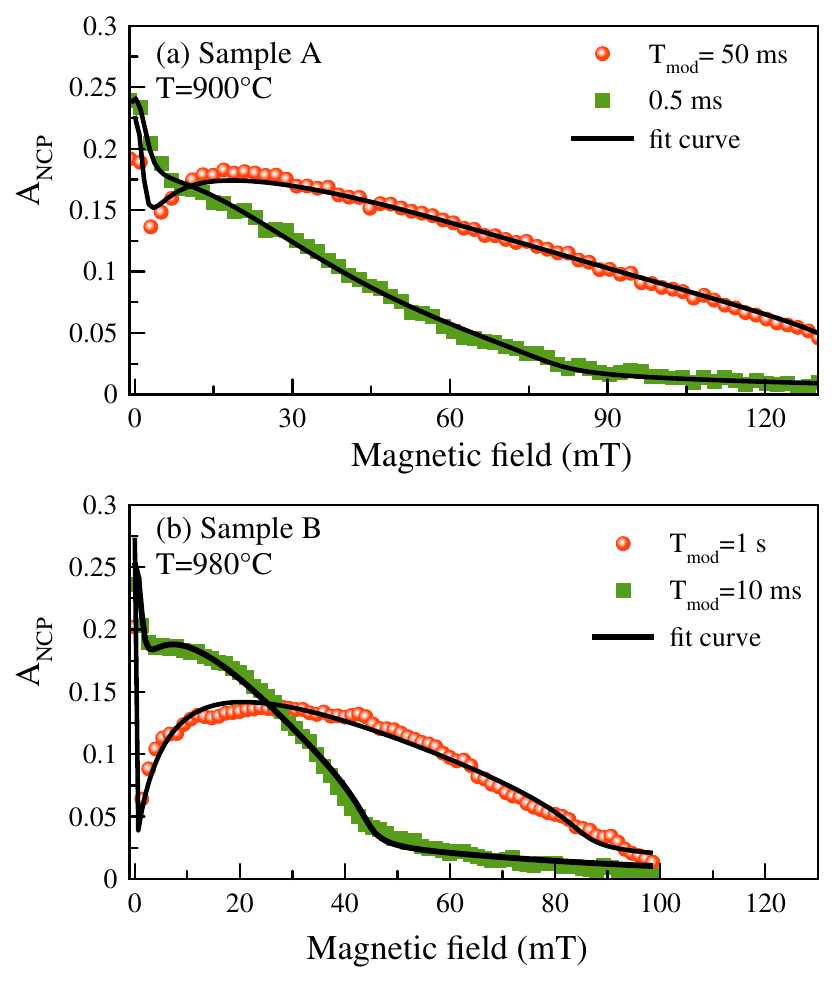}
\caption{(Color online) Examples of the Hanle curve simulations for samples A (a) and B (b) for different modulation periods given in the legends. Symbols are the experimental data and solid lines are the fits.%
}\label{fig:Fig3}
\end{figure}
To solve the problem we have first obtained approximate values of the parameters. For this purpose we fixed one parameter, $b_e = 4$~mT for sample A and $b_e = 2$~mT for sample B, and obtained other parameters by simple fitting procedure using Eqs.~\eqref{eq:Eq9}~and~\eqref{eq:Eq10}. Then we solved the total equation using the obtained values of the parameters as the initial ones and setting the limits for their possible variations. We found that there is only one root of the equation, which satisfies the physical conditions: $S_z$ is the real and positive quantity. 

Numerical solution of the equation for different magnetic fields allowed us to simulate Hanle curves by appropriate choice of the fitting parameters. We have ignored some asymmetry of Hanle curves observed experimentally (see Fig.~\ref{fig:Fig1}) and simulated only a part of each Hanle curve measures at $B > 0$. An analysis has shown that the fitting parameters are not noticeably changed when another part of Hanle curves is modeled.

\subsection{Analysis of Hanle curves\label{subsec:C}}

The phenomenological model developed above allowed us to describe  well the non-trivial shape of Hanle curves measured for both samples at different modulation periods. Example of the Hanle curves obtained in the model are shown in Fig.~\ref{fig:Fig3}. The good correspondence of the measured and simulated Hanle curves allows us to obtain values of the fitting parameters at each modulation period and, therefore, to evaluate their frequency dependence. Although there are several fitting parameters, values of most parameters can be determined independently because they control different features of Hanle curves. In particular, parameters $B_{Nd}$ and $B_{Nq}$, describing the photoinduced dipole and quadrupole nuclear spin polarization, determine the central part with W-structure and the peripheral part of Hanle curves, respectively. 

\begin{figure}[t]
\includegraphics[width=1\columnwidth,clip]{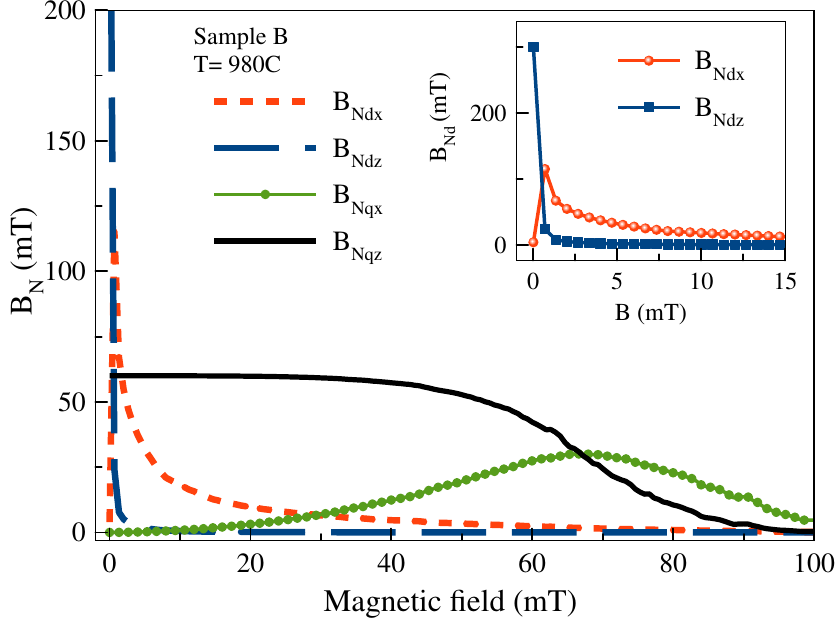}
\caption{(Color online) Examples of the magnetic field dependences of the longitudinal and transverse components of the dipole and quadrupole nuclear fields calculated for sample B using Eq.~\eqref{eq:Eq10}. The parameters used in the calculation are extracted from the Hanle curve measured at the modulation period $T_{mod}= 300$~ms. The magnetic field dependence of $B_e$ is taken from the experimentally measured Hanle curve. Inset shows behavior of components of the dipole field at small magnetic fields. At zero magnetic field, $B_{Ndz}=300~\mathrm{mT}$.%
}\label{fig:Fig4}
\end{figure}

Examples of magnetic field dependences of the dipole and quadrupole components of nuclear spin polarization are shown in Fig.~\ref{fig:Fig4}. As seen, the $x$- and $z$-component of the dipole field have large magnitude in small magnetic fields. In particular, the dipole component $B_{Ndz}$ has a maximal value at zero magnetic field and rapidly decreases with $B$ while component $B_{Ndx}$ rapidly rises in the same range of magnetic field (see insert in Fig.~\ref{fig:Fig4}).  As it is discussed in Ref.~\onlinecite{KuznetsovaPRB13}, such behavior of nuclear field is responsible for the W-structure in Hanle curves. Subsequent decrease of the $B_{Ndx}$ component completes the W-structure. Beyond the W-structure, i.e., in large magnetic fields, the dipole component of nuclear field is virtually absent.

The quadrupole field is weakly changed in small magnetic fields. In particular, $x$-component of the field is almost zero while $z$-component has some finite, almost constant, value. It is the component, which stabilizes the electron spin polarization making the Hanle curve broad at slow modulation of excitation polarization. At large external magnetic fields, the dipole field almost disappears and the wings of Hanle curve is mainly determined by competition of $x$- and $z$-components of the quadrupole field. As one can see in Fig.~\ref{fig:Fig4}, the $z$-component rapidly decreases at large $B$ and the $x$-component increases that results in relatively sharp decrease of electron spin polarization observed experimentally. So, the dipole field forms the W-structure and the quadrupole field forms the wings of Hanle curve.

Let us now discuss other parameters of the model.
Parameter $B_{\tau}$ is determined by time $\tau_{e}$ of the electron spin relaxation, see comment to Eq.~\eqref{eq:Eq6}. As it was mentioned above, $\tau_{e}$ depends on the excitation power but should be independent of the modulation period. Therefore we fixed its value, $B_{\tau} = 18$~mT.
This value is obtained from the Hanle curve width at the fastest modulation used when the nuclear spin effects are negligibly small. 

Parameter $b_e$ [see Eq.~\eqref{eq:Eq10}] characterizes the Knight field $B_e$ averaged over all the nuclei interacting with electron spin. The magnitude of this parameter is determined by the electron density on the nuclei~\cite{OO}. The described above simulations of the Hanle curves have shown that this parameter has to be changed under variation of the modulation period. In particular, $b_e = 5.3$~mT at slow modulation ($T_\mathrm{mod} > 0.01$~s) and $b_e = 8$~mT at fast modulation ($T_\mathrm{mod} < 0.01$~s) for sample A. For sample B, $b_e = 1.13$~mT at slow modulation ($T_\mathrm{mod} > 0.3$~s) and $b_e = 3.6$~mT at fast modulation ($T_\mathrm{mod} < 0.1$~s). We assume that this variation of $b_e$ with the modulation period is due to different rates of spin relaxation for different nuclear states. If the relaxation of some nuclear states is slower than the modulation period, such nuclear states are ''switched off'' from the joint electron-nuclear spin dynamics. Correspondingly, the Knight field should be averaged over a subset of nuclear states, which are not ''switched off''. Difference in the magnitudes of $b_e$ for sample A and B is explained by different electron densities on nuclei in these samples. Sample A contains QDs annealed at lower temperature ($T_\mathrm{ann}=900$~$^{\circ}$C) than the sample B ($T_\mathrm{ann}=980$~$^{\circ}$C) so that the indium content is larger, the electron localization volume is smaller, and the hyperfine interaction is stronger in sample A~\cite{PetrovPRB08}.

Parameter $k$ describing nonlinear splitting of the $\pm 3/2$ etc. doublets in magnetic field [see Eq.~\eqref{eq:Eq11b}] is found to be almost independent of modulation period for both  samples. Its average value is: $k = 0.9 \times 10^{-4}$~mT$^{-1}$ for sample A and $k = 1.8 \times 10^{-4}$~mT$^{-1}$ for sample B. The obtained values of $k$ can be compared with those found from Zeeman splittings of the $\pm 3/2$ states in different nuclei. According to the data of Ref.~\onlinecite{KuznetsovaPRB14}, $k$(Ga) = $20 \times 10^{-4}$~mT$^{-1}$, $k$(As) =$1.3 \times 10^{-4}$~mT$^{-1}$ for sample A and $k$(Ga) = $40 \times 10^{-4}$~mT$^{-1}$, $k$(As) = $7 \times 10^{-4}$~mT$^{-1}$ for sample B. As seen, these values considerably differ for the Ga and As nuclei and are larger than those obtained from the modeling of Hanle curves.

Possible reason for the difference of the $k$ values obtained from the fitting of experimental data and from the splittings can be related to the fast phase relaxation of nuclear spin polarization caused by fluctuating electron spin polarization under strong optical pumping used in the experiments. An analysis shows~\cite{KKavokin} that this relaxation should additionally weaken the effect of transverse magnetic field on the nuclear spin dynamics. 

Another possible reason is a contribution of the As nuclei in an asymmetric atomic configuration containing one or few In neighbors. Crystal field gradient caused by a statistical occupation of lattice nodes by the In and Ga atoms gives rise to a quadrupole splitting of spin states in the As nuclei~\cite{OO}. The principal axis of the gradient may be oriented along different crystal axes.  The quadrupole splitting in these nuclei is stronger, therefore the value of $k$ should be smaller. These nuclei can be responsible for stabilization of the electron spin polarization at large magnetic fields and, correspondingly, for the wings of Hanle curves observed experimentally. This contribution also explains the fact that the widths of Hanle curves for sample A and sample B do not strongly differ (see Fig.~\ref{fig:Fig1}) although the lattice deformation in sample A is three times larger compared to that in sample B~\cite{KuznetsovaPRB14}.

 Finally we should note, that the contribution of In nuclei into the effect of stabilization of the electron spin polarization is negligible because of wide spread of Zeeman splittins of different states ($m = \pm 3/2, \pm 3/2, \ldots, \pm 9/2$). 
\begin{figure}[t]
\includegraphics[width=1\columnwidth,clip]{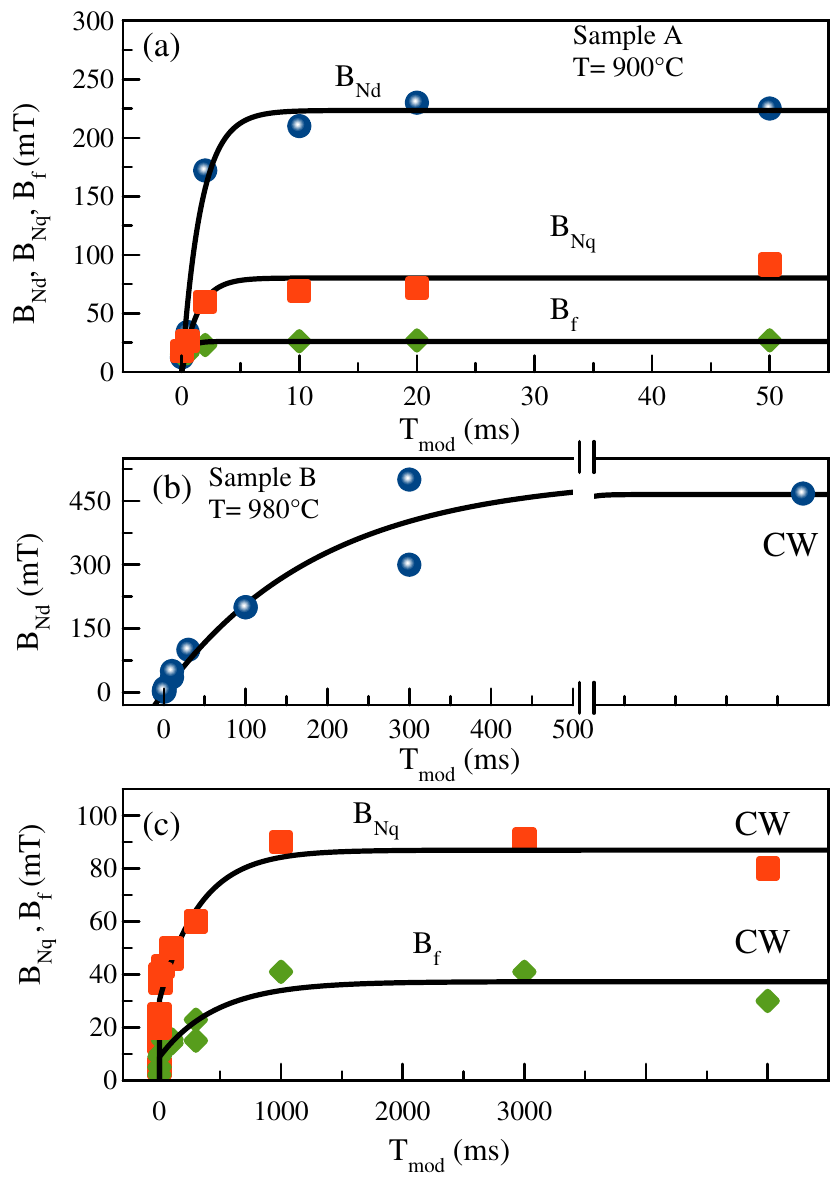}
\caption{(Color online) Dependences of the dipole field $B_{Nd}$, quadrupole field $B_{Nq}$, and the field of nuclear spin fluctuations $B_f$ on modulation period $T_\mathrm{mod}$ for sample A (a) and sample B (b, c). Symbols are the values extracted from the analysis of experimental data. Solid lines are the fits by Eqs.~\eqref{eq:Eq12} with characteristic times: for sample A: $\tau_{Nd} = 1.6$~ms, $\tau_{Nq} = 1.4$~ms, $\tau_{f} = 0.5$~ms; for sample B: $\tau_{Nd} = 191$~ms, $\tau_{1Nq} = 0.18$~ms, $\tau_{2Nq} =328$~ms, $\tau_{1f} = 0.2$~ms, $\tau_{2f} = 467$~ms.%
}\label{fig:Fig5}
\end{figure}
\subsection{Dynamics of nuclear fields}

The simulation of Hanle curves described above allows us to analyze evolution of the dipole and quadrupole nuclear fields at the modulation of excitation polarization. Figure~\ref{fig:Fig5} shows the evolution of initial (photoinduced) values of nuclear fields ${B}_{Nd}$ and ${B}_{Nq}$ for both the samples. The magnitudes of nuclear fields, in particular, of the dipole component, obtained in the simulations have relatively large spread. As it is already discussed (see Fig.~\ref{fig:Fig4}), the dipole component significantly differs from zero only at small magnetic fields in the range of W-structure of the Hanle curves. Therefore, any small inaccuracy of experimental data in this range noticeable affects the component. The quadrupole component is determined in the larger magnetic field range and, therefore, its magnitude is found with less uncertainty. Nevertheless, in spite of the spread, the obtained values of the dipole and quadrupole components demonstrate certain tendency in evolution of nuclear spin polarization.

As seen from the figure~\ref{fig:Fig5}, all the nuclear fields tend to go to some stationary values at slow enough modulation. These stationary values are very different for different nuclear fields and different samples. For example, as it is shown in Fig.~\ref{fig:Fig5}(a), the dipole field $B_{Nd}$ is only three times larger than the quadrupole field $B_{Nq}$ in sample A. For the strongly annealed sample B, the dipole field is at least of 10 times larger than the quadrupole field [compare Figs.~\ref{fig:Fig5}(b) and ~\ref{fig:Fig5}(c)]. So, the annealing slightly decreases the quadrupole field (which is very expectable effect) and drastically increases the dipole field achievable at the CW or slowly modulated excitation.

Both the dipole and quadrupole fields decrease with shortening the modulation period. The decrease of nuclear fields is naturally explained by some inertia of nuclear spin system, which does not allow it to be reoriented during the half-period of the modulation. This effect enables to estimate the characteristic relaxation times for each nuclear field. The dependences of the nuclear field amplitudes on the modulation periods for sample A can be approximated by simple phenomenological equation:
\begin{equation}
B_N=B_{N\infty}\left[1 - \exp\left(-\frac{T_\mathrm{mod}}{\tau_N}\right)\right],			
\label{eq:Eq12}					
\end{equation}
Similar equation well describes evolution of the dipole nuclear field for sample B. At the same time, evolution of the quadrupole nuclear field in this sample is not exponential and we have to use more complicate equation:
\begin{equation}
B_N=B_{N\infty}\left[1- a^2\exp\left(-\frac{T_\mathrm{mod}}{\tau_{N1}}\right)- b^2 \exp\left(-\frac{T_\mathrm{mod}}{\tau_{N2}}\right)\right],			
\label{eq:Eq13}				
\end{equation}
with condition $a^2 + b^2 = 1$. Here $B_{N\infty}$ is the value of nuclear field under the continuous wave excitation. 

As seen from Fig.~\ref{fig:Fig5}, the relaxation time of the dipole field for the stronger annealed sample B is larger by more than two orders of magnitude comparing to that for sample A. So drastic difference in the relaxation rates points out high sensitivity of the nuclear spin dynamics to quadrupole effects. We should mention also that the relaxation dynamics in bulk n-GaAs, where the quadrupole splitting is very small, is further slowed down by a few orders of magnitude~\cite{KalevichJTF82, RyzhovAPL15, RyzhovSciRep16, Kotur16}. 

Dynamics of quadrupole field in sample A is characterized by a relaxation time, which is close to that for dipole field in this sample. However, the dynamics in sample B is characterized by two relaxation times. The first one is close to that for sample A and the second one is only several times smaller than relaxation time of dipole field in this sample. Possible reason for such behavior of quadrupole field in sample B is discussed in the next section.

Fitting of the Hanle curves allowed us to obtain the effective field of nuclear spin fluctuations, $B_{fz}$. As one can see in Fig.~\ref{fig:Fig5}, the amplitude of fluctuations decreases with decreasing period of the modulation. The possible origin of this unexpected, at the first glance, effect is discussed in the next section. 
The dynamics of $B_{fz}$ is similar to dynamics of the quadrupole field and is characterized by a single relaxation time for sample A and two relaxation times for sample B.  

\section{Discussion}

The phenomenological model used in the previous section for analysis of the experimental data is based on approximation of the well separated nuclear spin doublets. This approximation is valid in some limited range of the transverse magnetic field when the Zeeman splitting of the doublets is considerably smaller than the quadrupole splitting. However the experimental data analyzed in the present work are measured in the relatively wide range of magnetic field of about $\pm 100$~mT where the Zeeman and quadrupole splittings become comparable, see Ref.~\onlinecite{KuznetsovaPRB14}. In such magnetic fields, the dipole ($\pm 1/2$) and quadrupole ($\pm 3/2$,...) states are mixed that makes consideration of the dipole and quadrupole fields in large magnetic fields to be not applicable. A more accurate microscopic model is required for analysis of the spin dynamics in quadrupole nuclei. To the best of our knowledge, there is no such model so far. Therefore, we consider the results obtained in the framework of our model as qualitative, rather than quantitative, characteristics of the nuclear spin system.

The most exciting experimental result is the drastic slowing down of nuclear spin relaxation at the increase of annealing temperature from $900$~$^{\circ}$C for sample A to $980$~$^{\circ}$C for sample B. This decrease of relaxation rate is directly seen in Fig.~\ref{fig:Fig1} and the model allows us to estimate the rate. In principle, the annealing increases the localization volume for the resident electrons that may result in a decrease of the relaxation rate via hyperfine coupling with electron spins~\cite{MaletinskyPRL07, WustNature16}. However, as it is pointed out in Ref.~\onlinecite{PetrovPRB08}, the annealing gives rise only to the two-fold increase of the volume and, correspondingly, to the two-fold decrease of the hyperfine interaction that certainly cannot explain so large difference in relaxation rates.  

Other result of our modeling is that the nuclear spin dynamics is non-trivial and consists of a fast process with characteristic time of order of $1$~ms and a slow process with characteristic time of about $1$~s (see Fig.~\ref{fig:Fig5}). We assume that the slow process is the relaxation of the dipole nuclear field in the conditions when dynamics of the dipole and quadrupole components is decoupled. 
These conditions is probably realized in the stronger annealed sample B where the dipole field is strongly magnified compared to the quadrupole field [see Figs.~\ref{fig:Fig5}(b) and (c)]. 

The dynamics of nuclear fields in sample A is much faster and characterized by practically the same relaxation rate (within experimental errors) for the dipole and quadrupole components [see Fig.~\ref{fig:Fig5}(a)].
Such dynamics in this sample is possibly caused by a mixing of the dipole and quadrupole nuclear spin states due to tilting of the principal axis of the electric field gradient tensor or presence of some biaxiality of the tensor~\cite{SokolovPRB16}. 

The relaxation of quadrupole field in sample B [Fig.~\ref{fig:Fig5}(c)] is characterized by the presence of fast and slow components. It is also possibly due to an asymmetry of the tensor. Another possible reason is that the quadrupole component of the nuclear field in this sample is small compared to the dipole one and even small error in the separation of the components during the experimental data processing may result in a noticeable admixture of the dipole component. Such admixture may be responsible for the complex shape time dependence for $B_{Nq}$ seen in Fig.~\ref{fig:Fig5}(c).

Large difference of relaxation times for the dipole and quadrupole components of nuclear field requires separate discussion. The slow relaxation in the $\pm 1/2$ nuclear spin system is inherent property of the system well known in the nuclear magnetic resonance~\cite{Abragam}.  In this process, the angular momentum $\pm 1$ should be transferred from the nucleus to a phonon. However there are no such phonons in the phonon bath. Hyperfine interaction with resident electrons accelerates this process~\cite{MaletinskyPRL07, WustNature16} but it still remains slow. 

Direct relaxation between the $\pm 3/2$ states is also ineffective. However any modulation of the electric field gradient should provoke an efficient relaxation $\pm 3/2 \rightarrow \pm 1/2$. In particular, the crystal field gradient can be modulated by fluctuations of the carrier density~\cite{DengPRB05, PagetPRB08, Kotur16}. In the case of QDs with relatively deep potential well for carriers, the fluctuations are small, at least at low sample temperature. However, in the case of optical excitation of QDs, the fluctuations may be much larger due to separate capture of electrons and holes so that this mechanism of relaxation of the quadrupole states may become effective. After the $\pm 3/2 \rightarrow \pm 1/2$ relaxation, rapid precession of the nuclear spins in the transverse magnetic field mixes the $\pm 1/2$ states due to large effective nuclear $g$ factor for this doublet. In particular, the precession frequency is of order $10^4$~Hz in magnetic field of 1~mT. The backward relaxation $\pm 1/2 \rightarrow \pm 3/2$, which occurs at any moment of the precession, should give rise to effective destruction of the quadrupole field. This simplified picture of the relaxation process may explain the rapid relaxation of quadrupole field observed experimentally. 

Finally we should mention about one more effect observed at the modulation of the excitation polarization. This is the suppression of the effective field of nuclear spin fluctuations, which is observed at the shortening of the modulation period [see Fig.~\ref{fig:Fig5}(c)]. The suppression unambiguously follows from the fact of the strong narrowing of the Hanle curve down to the purely electron peak at the fast enough modulation of polarization. We assume that the strong optical pumping with rapidly alternating polarization equalizes the population of nuclear states with spin down and spin up and, hence, suppresses the longitudinal component of nuclear spin fluctuations, which supports orientation of the electron spin. 
The transverse component of the field of nuclear spin fluctuations while, of course, remain, but, as the analysis showed, their contribution  to the formation of the Hanle curve is negligible small 
The mentioned above almost total coincidence of the dynamics of the nuclear spin fluctuation and of the quadrupole nuclear field suggests that the quadrupole states mainly contribute to $z$-component of the nuclear spin fluctuations. Similar conclusion has been made in Ref.~\onlinecite{Sinitsyn-PRL2012}.

\section{Conclusion}
Strong modification of Hanle curves observed under modulation of excitation polarization is demonstrated to contain valuable information about the dynamics of coupled electron-nuclear spin system in the studied (In,Ga)As/GaAs QDs. To extract this information, we have developed a simplified phenomenological model considering separate dynamics of the dipole and quadrupole nuclear fields. In particular, the quadrupole field can efficiently stabilize electron spin polarization in large magnetic fields up to $100$~mT. At the same time, the relatively fast relaxation of the quadrupole nuclear states to the $\pm 1/2$ states may considerably shorten the electron spin lifetime. In the studied samples with different quadrupole splittings, the lifetimes differ by more than two orders of magnitude.

\section*{Acknowledgments}
The authors thank K.V. Kavokin for fruitful discussion. The work is supported by the Russian Foundation for Basic Research and the Deutsche Forschungsgemeinschaft in the frame of International Collaborative Research Center TRR160 (Project No. 15-52-12020). I.V.I. acknowledges the support of the Russian Foundation
for Basic Research (Contract  No. 16-02-00245A) and SPbSU (Grant No. 11.38.213.2014).


\begin{thebibliography}{}
\bibitem{Note-field} Here we take into account that the electron $g$ factor is isotropic so that the electron "feels" the effective nuclear field, which direction coincides with that of nuclear spin polarization.
\end{thebibliography}
\end{document}